\documentclass[10pt]{article}
\usepackage{times}
\usepackage{fullpage}
\usepackage{amsfonts}
\usepackage{amsmath}
\usepackage{latexsym}
\usepackage{amssymb}
\usepackage{bbm}
\usepackage[pdftex]{graphicx}

\newcommand{\captionfonts}{\small}
\newcommand{\N}{\mathbb{N}}
\newcommand{\E}{\mathbbm{E}}

\makeatletter  
\long\def\@makecaption#1#2{%
\vskip\abovecaptionskip
\sbox\@tempboxa{{\captionfonts #1: #2}}%
\ifdim \wd\@tempboxa >\hsize
{\captionfonts #1: #2\par}
\else
\hbox to\hsize{\hfil\box\@tempboxa\hfil}%
\fi
\vskip\belowcaptionskip}
\makeatother   

\begin{document}
\title{Fixed-points in Random Boolean Networks: \\  The impact of  parallelism in the scale-free topology  case
\footnote{Partially supported by Programs ``Instituto Milenio ICDB'',
Basal-CMM (A.O., I.R.), Fondecyt 1090156 (I.R.), and Fondecyt 1090290 (A.O.).}}

\author{\normalsize{Pablo Moisset de Espan\'{e}s$^1$, Axel Osses$^{2,3}$ and Ivan Rapaport$^{2,3}$} \\
{}\\
\small{\llap{$^1$}}Instituto de Din\'amica Celular y Biotecnolog\'{\i}a, Universidad de Chile\\
\small{\llap{$^2$}}Departamento de Ingenier\'{\i}a Matem\'atica, Universidad de Chile\\
\small{\llap{$^3$}}Centro de Modelamiento Matem\'atico (UMI 2807 CNRS), Universidad de Chile}



\date{}
\maketitle
\thispagestyle{empty}

\begin{abstract}
Fixed points are fundamental states in any dynamical system. In the case
of gene regulatory networks (GRNs) they correspond to stable genes
profiles associated to the various cell types. We use Kauffman's
approach to model GRNs with random Boolean networks (RBNs). We start
this paper by proving that, if we fix the values of the source nodes
(nodes with in-degree 0), the expected number of fixed points of any RBN
is one (independently of the topology we choose). For finding such fixed
points we use the $\alpha$-asynchronous dynamics (where every node is
updated independently with probability $0<\alpha<1$).  In fact, it is 
well-known that asynchrony avoids the cycle attractors into which parallel
dynamics tends to fall. We perform simulations and we show the
remarkable property that, if for a given RBN with scale-free topology 
and $\alpha$-asynchronous dynamics an initial configuration reaches a
fixed point, then every configuration also reaches a fixed point. By
contrast, in the parallel regime, the percentage of initial
configurations reaching a fixed point (for the same networks) is
dramatically smaller.  We contrast the results of the simulations on
scale-free  networks with the classical Erd\"{o}s-R\'enyi model of 
random networks.  Everything indicates that scale-free networks are
extremely robust. Finally, we study the mean and  maximum time/work 
needed to reach a fixed point when starting from randomly chosen initial
configurations.
\end{abstract}

\section{Introduction}

A {\em Random Boolean Network} (RBN) is a model of a GRN. This model was
introduced by Kauffman in 1969~\cite{Kauffman1969} and   it corresponds
to a directed graph composed by $N$ genes (nodes) where each of these
genes can be either expressed (state 1) or not expressed (state 0). Each
gene $v$ receives  $K$ randomly chosen genes as input ($K$ nodes
pointing to $v$). In other words, the dynamics is defined
\emph{locally}. The idea of Kauffman was to choose,  independently for
each node $v$, any of the $2^{2^K}$ possible functions $f_v:\{0,1\}^K
\to \{0,1\}$  with equal probability. Finally, a global dynamics
$f:\{0,1\}^N \to \{0,1\}^N$ was defined by applying the local functions
\emph{in parallel} (sometimes this is referred to as \emph{synchronous
updates}). Kauffman found, through simulations, the existence of a phase
transition at $K=K_c=2$ (from an ordered phase to a chaotic phase). He
also suggested that the number of attractors grows exponentially with
$N$ in the chaotic phase, proportional to $N$ in the ordered phase, and 
proportional to $\sqrt{N}$ at the critical value $K_c=2$.

Kauffman's model (and some natural generalizations) motivated a lot of
work carried out mainly by physicists. They tackled problems arising from
Kauffman's definitions both analytically and numerically: the critical
in-degree value~\cite{Bastolla1996,Derrida1986}, the number of
attractors~\cite{Bilke2001,Derrida1986}, the length distribution of the
attractors~\cite{Aldana2003,Bhattacharjya1996}, etc.

From the biologists point of view a fundamental challenge of  the
post-genomic era is the possibility of simulating the dynamics of
\emph{real} genetic networks. The enormous amount of  available data of
molecular interactions within the cell made it possible to examine
critically the original RBN model. The first observation was that the
topology assumption was inadequate. Contrasting the uniform topology
assumed in the original RBNs, it has been shown that real genetic
networks exhibit a \emph{scale-free} 
topology~\cite{Albert2005,Christensen2007,Fox2001,Oosawa2002}. In such
topologies a small fraction of the genes are highly connected whereas
the majority of the genes are poorly
connected~\cite{Albert2002,Strogatz2001}. Therefore, RBNs dynamics with
scale-free  topology started to be intensively
studied~\cite{Aldana2003-1,Damiani2008,Iguchi2007}.

Another criticism towards Kauffman's model was that nodes were updated in
parallel.  Experimental results confirmed a rather intuitive fact: that
genes transition between expressed and non-expressed states at different
times~\cite{Davidson2002,Edwards2006}. Informally, there is no global
clock that allows transition to happen only at ticks. Therefore, RBNs with
scale-free topology and asynchronous dynamics is a natural model to be
analyzed~\cite{Darabos2007}. We would also like to point out that
asynchrony in the classical RBN model has also  been studied
in~\cite{Gershenson2004,Harvey1997,Mesot2003} and~\cite{Schonfisch1999}.

\bigskip

{\bf Our main contributions.} \emph{Fixed points}  are fundamental states in
any dynamical system. In the case of GRNs they correspond
to stable gene expression profiles associated  to the various cell
types. This interpretation has been used for modeling  europhil
differentiation~\cite{Huang2005},  expression patterns of the segment
polarity genes in  {\emph{Drosophila melanogaster}}~\cite{Albert2003},
flower organ specification in {\emph{Arabidopsis
thaliana}}~\cite{Alvarez2008}, etc.

The goal of this paper is to study the of existence fixed points in RBNs
with scale-free topology. Notice that a fixed point does not depend on
the timing (synchronous/asynchronous)  of the update rule.  In fact, in
a fixed point every gene is in a stable state with  respect to its local
input. Therefore, there is no way to change such global configuration.
In that sense, a fixed point is a very robust object of a network,
despite the fact that its basin of attraction can change depending on
how the update rule is implemented.

Once the topology of the network is (randomly) generated some nodes will
have only outgoing  arcs. These were called \emph{source nodes}  by
Albert~\cite{Albert2005} and we will use the same terminology here.
Since their states do not depend on the state of any other node we fix
their values arbitrarily. In this paper we prove that, for every given
assignment to the input nodes, the expected number of fixed points is
exactly one (for \emph{every} topology).

Now the main (and natural) question arises.  Given a RBN, how can we
actually \emph{find} its fixed points? It is clear that testing all the
$2^N$ configurations is impossible. For answering the question we come
back to the issue of asynchrony.   In fact, in 1994 Bersini and Detours
studied an asynchronous version of the cellular automaton
Game-of-Life~\cite{Bersini1994}. They observed that the introduction of
asynchrony modified the dynamics from a behavior with \emph{long
transients} to a behavior with fixed points. This is rather intuitive: 
asynchrony is a way to avoid the cycle attractors the deterministic
(parallel) implementation tend to fall into.

Therefore, the strategy is clear. Given a RBN,  we implement the
$\alpha$-asynchronous dynamics for different values of $\alpha$ ($0 \leq 
\alpha \leq 1$) ~\cite{Fates2006}.  Roughly, this means that, at each
time step,  each gene is updated independently  with probability
$\alpha$.  When $\alpha$ varies from 1 down to 0 the dynamics evolves
from the fully deterministic synchronous regime  to a more asynchronous
regime. When $\alpha=0$ we choose randomly only one node at each step.

Our simulations show that RBNs with scale-free topology  for which there
exist a fixed points \emph{every} initial configuration converges to a
fixed point when $0 \leq \alpha < 1$. On the other hand, when
$\alpha=1$, the   percentage of initial configurations that reached a
fixed point varies greatly form one network to another. In some cases
the percentage is close to 0. In average (considering all the networks
we use) the percentage is $\sim$28.9\%.

In order to find properties which could be associated exclusively to the
topology of the network we compare the results of the simulations on
scale-free  networks with the results on the classical Erd\"{o}s-R\'enyi
model of  random networks~\cite{Erdos1959}. The main difference with 
the scale-free topology is that here we generated networks for which the
percentage of initial configurations converging to a fixed point is
close to $0$. Finally, we study the mean and  maximum time/work 
needed to reach a fixed point when starting from randomly chosen initial
configurations.

Therefore, a remarkable and distinguishable dynamical property arise on
RBNs with scale-free topology: robustness of convergence under
asynchronous update. This fact could provide some insight about why such
topologies are ubiquitous in GRN. Furthermore, asynchronous updating
could be a natural mechanism present in GRNs in order to avoid cyclic
dynamics~\cite{Schonfisch1999}.

\section{Network model}
\label{sec:model}

A {\em Random Boolean Network (RBN)}  corresponds to a directed graph
composed by $N$ genes (nodes) where each of these genes can be either
expressed (state 1) or not expressed (state 0). We will refer to the
nodes of a RBN  as $v_1,v_2,\ldots,v_N$.  We define $K_i^{in}$ as the
in-degree of node $v_i$ and $K_i^{out}$ as the out-degree of node $v_i$.
Every zero in-degree  node is called a {\em source node}, while  every 
non-zero in-degree node is called an {\em internal node}. A {\em
configuration} of the network is a vector ${\mathbf s} \in \{0,1\}^N$ 
that associates a binary state to each of the nodes.

\subsection{Dynamics}
\label{subsec:dynamics}

We assign to each gene $v_i$  a  {\em local transition rule} $\phi_i:
\{0,1\}^{ K_i^{in}+1}\rightarrow \{0,1\}$. Informally, the value of
$\phi_i$ depends on the state of the $K_i^{in}$ input nodes together
with the state of $v_i$ itself. source nodes always remain in the same
state. More precisely, if $K^{in}_i=0$, then  $\phi_i(0)=0$ and
$\phi_i(1)=1$. For each internal node $v_i$ we construct randomly its
local transition function as follows.   Call $k$ the in-degree of $v_i$
(i.e, $k=K^{in}_i$). There are $2^{2^k}$ possible functions of the form
$f:\{0,1\}^k\rightarrow\{0,1\}$. A straightforward approach is to choose
one of these functions from a uniform probability distribution. 
Nevertheless, before selecting a function, we should rule out those
which do not strictly depend on all of its arguments (otherwise we would
not be respecting the network topology). To define this concept
precisely, we say that  a function $f:\{0,1\}^k\rightarrow\{0,1\}$ {\em
strictly depends on its arguments} iff for all $j\in\{1,2,\cdots,k\}$,
there exist $x_1,x_2,\cdots,x_{j-1},x_{j+1},\cdots,x_k \in \{0,1\}$ such
that  $f(x_1,x_2,\cdots,x_{j-1},0,x_{j+1},\cdots,x_k) \not =
f(x_1,x_2,\cdots,x_{j-1},1,x_{j+1},\cdots,x_k)$. We fix function
$\phi_i$ by selecting one (randomly) among all  those functions strictly
depending on all of its arguments.

The global update rule is characterized by a real parameter $\alpha \in
[0,1]$. We denote the  global transition rule by
$\mathbf{\Phi}_{\alpha}: \{0,1\}^N \rightarrow \{0,1\}^N$, and we define
it through the following protocol:

\begin{enumerate}

\item {\bf Select}  each internal node independently with probability
$\alpha$. We call {\em selected nodes} this set of randomly chosen
internal nodes. For the special case  $\alpha=0$ we select  randomly 
{\em a single} internal node (i.e, the set of selected nodes is a
singleton).

\item {\bf Update} in parallel all the selected nodes (applying the
local transition rule in all the nodes belonging to such set). Do not
change the state of the other nodes (input nodes and non-selected
nodes).

\end{enumerate}

Let ${\mathbf s}_t \in \{0,1\}^N$ be a configuration of a RBN at time $t
\in \N$. A {\em stochastic trajectory}, starting from the initial
configuration ${\mathbf s}_0$, is the sequence ${\mathbf s_0},{\mathbf
s_1},{\mathbf s_2},\ldots$, where  ${\mathbf
s_{i}}={\mathbf{\Phi}}_{\alpha}({\mathbf s_{i-1}})$.

The parameter $\alpha$ can be thought of as a measure of parallelism in
the update process. The strictly sequential-random policy rule is
captured by $\alpha=0$, where only one internal node is updated at each
step. Similarly, by making $\alpha=1$, we represent the full
parallel-deterministic policy rule, where  all the internal nodes are
updated at each step. If $\alpha=1$ we call the resulting trajectory the
{\em deterministic trajectory} of the system.  A configuration
$\mathbf{s}$ is a {\em fixed point} of a RBN iff
$\Pr\{\mathbf{\Phi}_{\alpha}(\mathbf{s})=\mathbf{s}$\}=1. This is
equivalent to say that $\mathbf{\Phi}_{1}(\mathbf{s})=\mathbf{s}$.
Therefore,   $\mathbf{s}$ is a fixed point  {\em regardless of the
choice of $\alpha$}.

We measure time simply by counting the applications of
$\mathbf{\Phi}_{\alpha}$. Therefore, the time to go from ${\mathbf s_0}$
to ${\mathbf s_T}$ in a trajectory is $T$. This notion of time neglects
the fact that the computational effort to evaluate
$\mathbf{\Phi}_{\alpha}$ depends on $\alpha$. The expected number of
$\phi$'s to be evaluated is $\alpha N_I$ where $N_I$ is the number of
internal nodes. Thus, we define the {\em work} to go from state
${\mathbf s_0}$ to ${\mathbf s_T}$ to be $\alpha T$ for all $\alpha>0$.
For the special case $\alpha=0$ we define the work as $\frac{1}{N_I}T$.

When considering the dynamics of the network with $\alpha=1$, it is
clear that the trajectory will eventually become cyclic. Note that if
the system reaches a fixed point, the length of the period is 1. Both
the time to enter the cycle and the period are bounded by $2^{N_I}$,
where $N_I$ is the number of internal nodes. However, if $\alpha < 1$,
the idea of cycle is insufficient to describe a trajectory that does not
reach a fixed point. Instead, we have a set of configurations which are
visited infinitely often. This greatly complicates the numerical
determination of any eventual convergence to a fixed point. We therefore
select a somewhat arbitrary time horizon to interrupt our simulations. 

Notice that, given an initial configuration $\mathbf{s}_0$, the set of
configurations reachable from $\mathbf{s}_0$ by repeated applications of
$\mathbf{\Phi}_{1}$ is a subset of the set of configurations potentially
reachable from $\mathbf{s}_0$ by repeated applications of
$\mathbf{\Phi}_{\alpha}$ when $\alpha<1$. In other words, the stochastic
trajectory can visit a larger portion of the state space than the
deterministic one.

A remarkable feature of selecting the local update functions randomly,
as we did, is that we can predict, in a statistical sense, the number of
fixed points in the network. Consider an arbitrary configuration
$\mathbf{s} \in \{0,1\}^N$. What is the probability of $\mathbf{s}$ to
be a fixed-point? First notice that  $\mathbf{s}$ is a fixed point if
and only if  $\mathbf{s}_i =
\phi_i({\mathbf{s}}_{i_1},\ldots,{\mathbf{s}}_{i_{k_i}})$ for every $i
\in \{1,\ldots,N\}$,  where  $v_{i_1},\ldots,v_{i_{k_i}}$ are the input
nodes of $v_i$. 

It follows from the definition of the $\phi_i$'s functions,  that
$\Pr\{\phi_i({\mathbf{s}}_{i_1},\ldots,{\mathbf{s}}_{i_k})=0\}=\Pr\{\phi_i({\mathbf{s}}_{i_1},\ldots,{\mathbf{s}}_{i_k})=1\}=\frac{1}{2}$
for every internal node $v_i$.  The reason for that is the following: if
a function $\phi_i$ strictly depends on all of its arguments then the
complementary function (the one where we replace 1's with 0's and 0's
with 1«s) also does. On the other hand,
$\Pr\{\phi_i({\mathbf{s}}_i)={\mathbf{s}}_i\}=1$ for every source node
$v_i$. Therefore, the probability of $\mathbf{s}$ to be a fixed-point is
$2^{-{N_I}}$, where $N_I$ is the number of internal nodes. 

Let $X_{\mathbf{s}}$ be the Boolean random variable that  equals 1 if
the configuration $\mathbf{s}$ is a fixed point and 0 otherwise.  The
expected number of fixed points is

$$\sum_{\mathbf{s} \in \{0,1\}^N}{\E(X_{\mathbf{s}})} = \sum_{\mathbf{s} \in \{0,1\}^N}2^{-{N_I}}=2^{N-N_I}$$

Note that $N-N_I=N_E$ is the total number of  source nodes of the
network. If we fix all these source nodes to some arbitrary value in the
initial configuration, then the expected number of fixed points goes
down to 1. In fact, in that case, the number of effective  different
configurations is $2^{N-N_E}=2^{N_I}$ and the expected number of fixed
points becomes $2^{N_I-N_I}=1$ instead of the original $2^{N-N_I}$.

Until this point, neither the definitions nor the theorem we just proved
make assumptions about the topology of the RBN. In the following
subsection we will describe two families of topologies that have been
studied in the literature.

\subsection{Topology}
\label{ssec:topology}

We start describing here a process by which we  construct  a
{\emph{directed scale-free network}} with $N$ nodes and average
in/out-degree equal to $k$. Let $N_0$ and $k$ be positive  integers such
that $k \leq N_0 \leq N$. The process starts from a directed clique with 
$N_0$ nodes (i.e, $N_0(N_0-1)$ arcs). Call these nodes $v_1,
v_2,\ldots,v_{N_0}$. The process now involves $N-N_0$ growth stages,
numbered $N_0+1,N_0+2,\ldots,N$. At each stage, a single node is added
to the network.

Call $v_i$ the node added at stage $i$. We will also add  $k$ edges to
the growing network. We toss a fair coin and proceed as follows:

\begin{enumerate}

\item
In the case of heads  we add $k$ edges pointing from $k$ different nodes
in $\{v_1,\ldots v_{i-1}\}$ towards $v_i$. These $k$ nodes are selected 
randomly following a {\em preferential attachment rule} such that the
probability of $v_j$ to be selected is proportional to $K^{out}_j+1$.

\item In the case of tails we add $k$ edges pointing from $v_i$ to $k$
different nodes in $\{v_1,\ldots v_{i-1}\}$. These $k$ nodes are
selected  randomly following a {\em preferential attachment rule}, such
that the probability of $v_j$ to be selected is proportional to
$K^{in}_j+1$.

\end{enumerate}


We will refer to the process just described as the {\em BA algorithm}
because it is based on previous work by Barab\'asi and Albert
~\cite{Barabasi1999}. Following~\cite{Iguchi2007}, we started with  a
clique of size $N_0=5$ and average in/out-degree $k=2$ to create the
topology using the BA method. If a RBN has a topology created by the BA
algorithm, we will call it a {\em scale-free network}.

The BA method, as defined here, never creates an edge from some node to
itself. The same assumption is present in~\cite{Barabasi1999} and~\cite{Iguchi2007} and is pervasive in the
literature. It is noteworthy that the theorem proved in
Subsection~\ref{subsec:dynamics} does not require a special topology.
Therefore, it still applies to networks where auto-regulation is
present. Thus, for given values for the input nodes, the expected number
of fixed points would still be 1. It is conceivable that the probability
distribution of the number of fixed points will change, though.

In the next sections, we compare the results on scale-free networks with
the classical Erd\"{o}s-R\'enyi model of random networks~\cite{Erdos1959}
(random directed graphs). By doing so we intend to find properties which
could be associated exclusively to the topology of the network. Random
networks of this well-known Erd\"{o}s-R\'enyi family can be easily 
generated.  Let  $p$ be a real number in the $[0,1]$ interval. Each
potential link $(v_i,v_j)$ is selected independently with probability
$p$. Therefore, the expected number of links in the Erd\"{o}s-R\'enyi
network is $pn(n-1)$, where $n$ is the number of nodes in the
Erd\"{o}s-R\'enyi network.  We will call this process the {\em ER
algorithm}. If a RBN has a topology created by the ER algorithm, we will
call it an {\em Erd\"{o}s-R\'enyi network}.


Conceptually, to make such comparison ``fair," we have to adjust the
parameters used by the generation algorithm so the networks generated
have some common statistical properties. It seems reasonable to preserve
both the number of internal nodes and the average in-degree. The first
parameter determines the number of states the system can be in, after
the state of the source nodes have been fixed. The average in-degree
is a measure of connectivity and density. Formally, if we run the
topology generation BA algorithm presented  for scale free networks with
parameter $N$ (recall that $N_0=5$ and $k=2$) we will obtain a
scale-free network $G_{BA}$. The average in-degree is $k=2$. Call $N_I$
the expected number of internal nodes of $G_{BA}$. Similarly, if we run
the topology generation ER algorithm with parameters $n$ and $p$, we
will obtain a graph $G_{ER}$.

The problem we wish to solve is: Given $N_I$ find $n$ and $p$ such that
the expected in-degree of nodes in $G_{ER}$ is 2 and the expected number
of internal nodes of $G_{ER}$ is $N_{I}$. Therefore,

\begin{align}
p\cdot(n-1)&= 2      \label{eqn:cond1}\\
n\cdot(1-(1-p)^{n-1}) &= N_I  \label{eqn:cond2}
\end{align}

From Eqn~\ref{eqn:cond1}, $p=\frac{2}{(n-1)}$. Substituting in
Eqn~\ref{eqn:cond2}:

\begin{align}
n\cdot\left(1-\left(1-\frac{2}{(n-1)}\right)^{n-1}\right)= N_I
\label{eqn:hard}
\end{align}

The only unknown is $n$, and although this equation is hard to solve
in closed form, the right hand side behaves linearly in the asymptotic
sense. Using simple calculus techniques, we can estimate the solution
to Eq~\ref{eqn:hard} as:

\begin{align}
n = \left(N_I-2\cdot e^{-2}\right)\frac{1}{1-e^{-2}} \label{eqn:easy}
\end{align}

The asymptotic approximation is so good that the rounding of $n$ to an
integer is the biggest source of error even for small values, say $10$,
of $N_I$. Therefore, for practical purposes, Eq~\ref{eqn:easy} gives the
exact answer. To use this formula you have to know $N_I$, though. In
spite of $N_I$ being determined by $N$, it is not straightforward to
find out an explicit formula. To obtain an approximation, we can simply
generate a suitable number of networks using the BA algorithm and
estimate $N_I$ as the average of the number internal nodes over all the
generated graphs.

\section{Simulations}
\label{sec:experiments}

We programmed a simulator using about 700 lines of portable ANSI C. We
ran a number of pseudo-random experiments. The main goal was to study the
influence of the parameter $\alpha$ in the dynamics of the networks. More precisely,
we were interested in answering,  {\emph{for a given network}}, the following questions
as a function of $\alpha$:

\begin{enumerate}

\item Are there  fixed points?

\item If yes,

\begin{enumerate}

\item how many?

\item what fraction of trajectories converge to a fixed point?

\end{enumerate}

\item If we restrict the analysis to those trajectories that converged to a fixed point,

\begin{enumerate}

\item what is the average time (number of iterations) until a fixed point is reached?

\item what is the average work  (total number of operations) 
\footnote{Notice that,
for $\alpha=0$, the amount of work per iteration is minimal (only 1
node is updated). On the other extreme, when $\alpha=1$, the amount of
work per iteration  is maximal (all the internal nodes are updated in parallel in each
iteration).} until a fixed point is reached?

\end{enumerate}

\end{enumerate}

We setup the time horizon to 50000 iterations. This number seems to
yield a robust determination of whether the network eventually reaches a
fixed point or not. For the scale-free networks we ran the BA algorithm
using parameters $N=100$ and $k=2$. We generated $50$ networks. For each
network we generated $1000$ initial configurations. The states were
generated randomly, but the values corresponding to source nodes were
set to zero. More precisely, to generate the initial configuration 
$\mathbf{s}$ of a network $N$, if $K^{in}_i=0$ then the $i$-th component
of $\mathbf{s}$ was set to zero. Otherwise, the $i$-th component of
$\mathbf{s}$ was generated randomly by tossing a fair coin.

Each one of the $50 \times 1000$ pairs network/initial configuration was
used as a starting point for 11 simulations, each one using a different
value of $\alpha$. The values of $\alpha$ were $0, 0.1,..., 0.9, 1$.
We ran each one of the $50\times 1000\times 11$ simulations until the
trajectory converged to fixed point or the time horizon was reached.

For the networks using the Erd\"{o}s-R\'enyi topology we had to compute the
input parameters for the ER algorithm ($n$ and $p$). To estimate
$N_I$, we generated $10000$ graphs using the BA algorithm, with $k=2$
and $N=100$. Then we divided the total number of internal nodes by
10000, which yielded $N_I=62.745$. By using Eq~\ref{eqn:easy}, we
determined $n=72$. From Eq~\ref{eqn:cond1} we obtained $p=0.028196$.

\begin{figure}[ht]
\centering
\includegraphics[height=55mm]{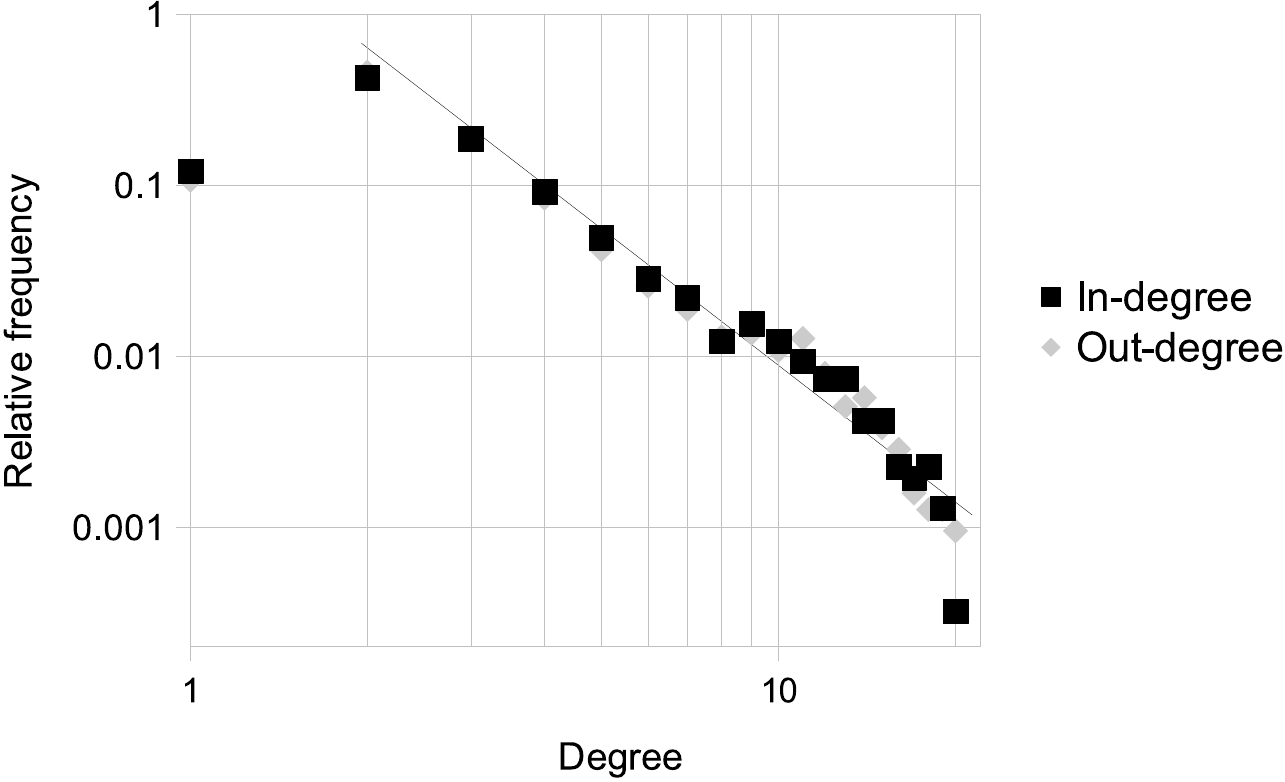}
\caption{Relative frequency of vertex degrees in the 50 scale-free networks}
\label{fig:degree_dist}
\end{figure}

With the two parameters, we repeated the process we used for the
scale-free networks: We created $50$ random networks and, for each one,
we generated $1000$ initial conditions. We also used the same values for
$\alpha$ and the same time horizon for the simulations. We verified that
the BA algorithm indeed produces graphs with vertex degrees that follow
a power law probability distribution. Figure~\ref{fig:degree_dist} shows
the relative frequency of each degree, for the 50 scale-free networks.
The diagram, in log-log coordinates, shows a high similarity with a
straight line. This is what we expect from a power-law distribution. The
noise for the higher vertex degrees is not surprising, as highly connected
vertices are rare and hence the population size is small.

\section{Results and analysis}

\subsection{Number of fixed points}

\begin{figure}[ht]
\begin{minipage}[t]{0.45\linewidth}
\centering
\includegraphics[height=55mm]{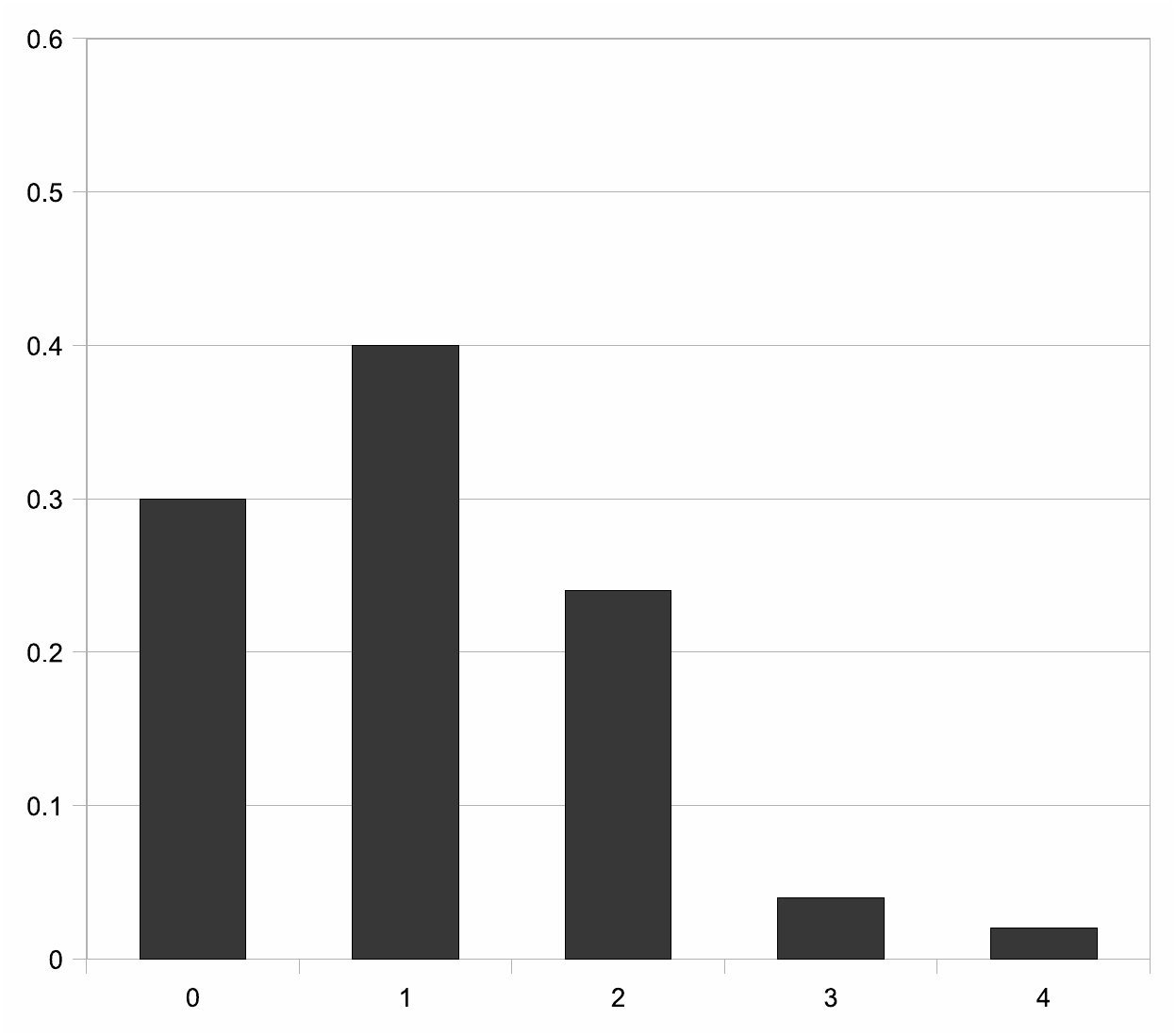}
\caption{Frequency of number of fixed points for scale-free networks.}
\label{fig:ba_effective_fixed_points}
\end{minipage}
\hspace{0.3cm}
\begin{minipage}[t]{0.45\linewidth}
\centering
\includegraphics[height=55mm]{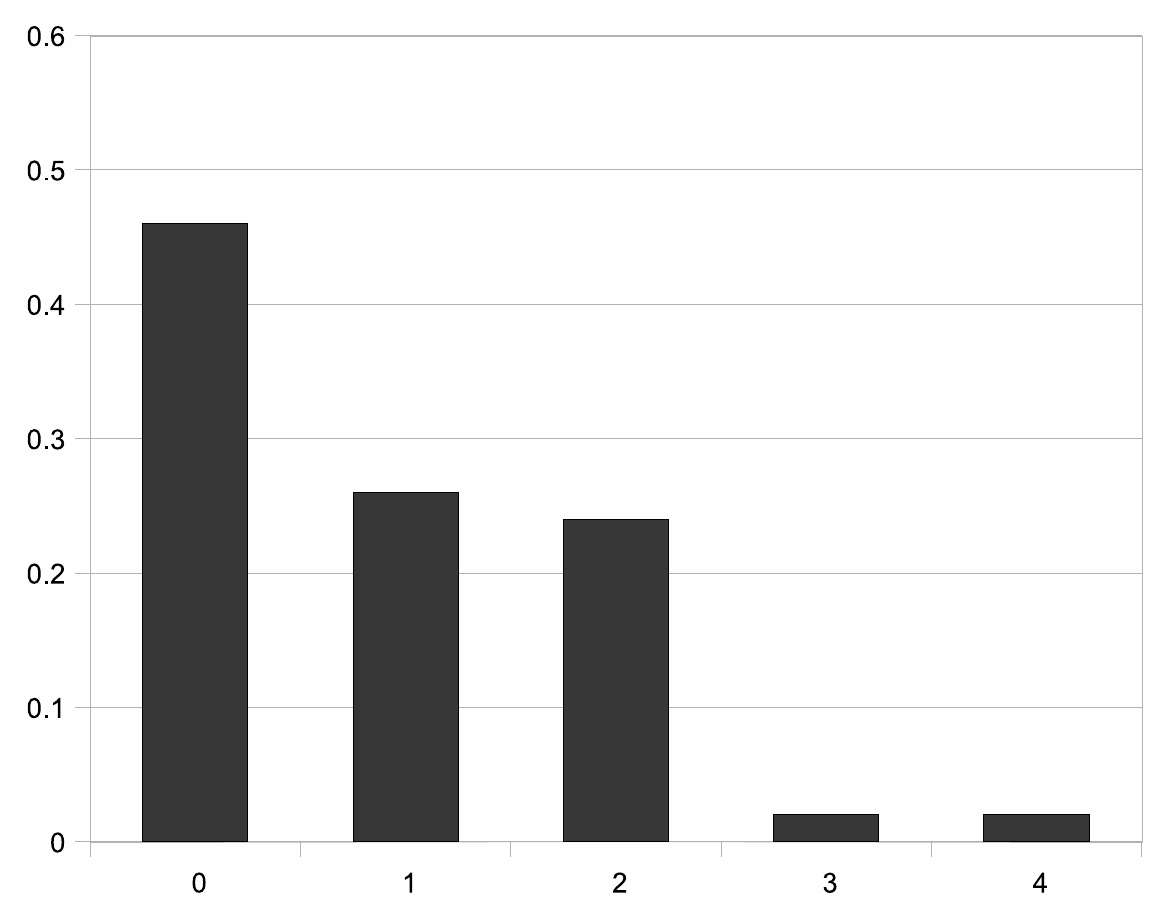}
\caption{Frequency of number of fixed points for Erd\"{o}s-R\'enyi networks.}
\label{fig:er_effective_fixed_points}
\end{minipage}
\end{figure}

For both the scale-free and the Erd\"{o}s-R\'enyi families we computed the
number of fixed points found per each network and we summarize the
results in Figures~\ref{fig:ba_effective_fixed_points}
and~\ref{fig:er_effective_fixed_points}. The vertical axis of the
histograms represent relative frequency over the 50 networks tested.
Note that in 15 out of the 50 generated scale-free networks, no fixed
point was found. More precisely, for all $\alpha$, none of the 1000 trajectories
was absorbed into a fixed point within the 50000 iterations used as
time horizon. The number of Erd\"{o}s-R\'enyi networks for which no
fixed-point was found is 23.

We proved analytically in Section~\ref{sec:model} that the expected
number of fixed points for any network is $1$. Unfortunately, the actual
distribution is hard to compute and therefore we use histograms as a
base for a numerical approximation. We can see in
Figure~\ref{fig:ba_effective_fixed_points} that for scale-free network 
the most likely number of fixed points is $1$. If we use the relative
frequencies to approximate probabilities, then the estimated expected
number of fixed points is $1.08$, which is close to the theoretical
prediction.

By contrast, for the Erd\"{o}s-R\'enyi model, we can see in
Figure~\ref{fig:er_effective_fixed_points} that the distribution gets
more skewed and $0$ becomes the most likely value. The estimation of
expected number of fixed points in this case yielded $0.88$. This estimation is
close to $1$ if we consider we are using only 50 networks and the number of
potential fixed points is about $2^{N_I}$, with $N_I\approx 63$, as we described
in Subsection~\ref{ssec:topology}.

\subsection{Fraction of converging trajectories}

\begin{figure}[ht]
\begin{minipage}[t]{0.45\linewidth}
\centering
\includegraphics[height=55mm]{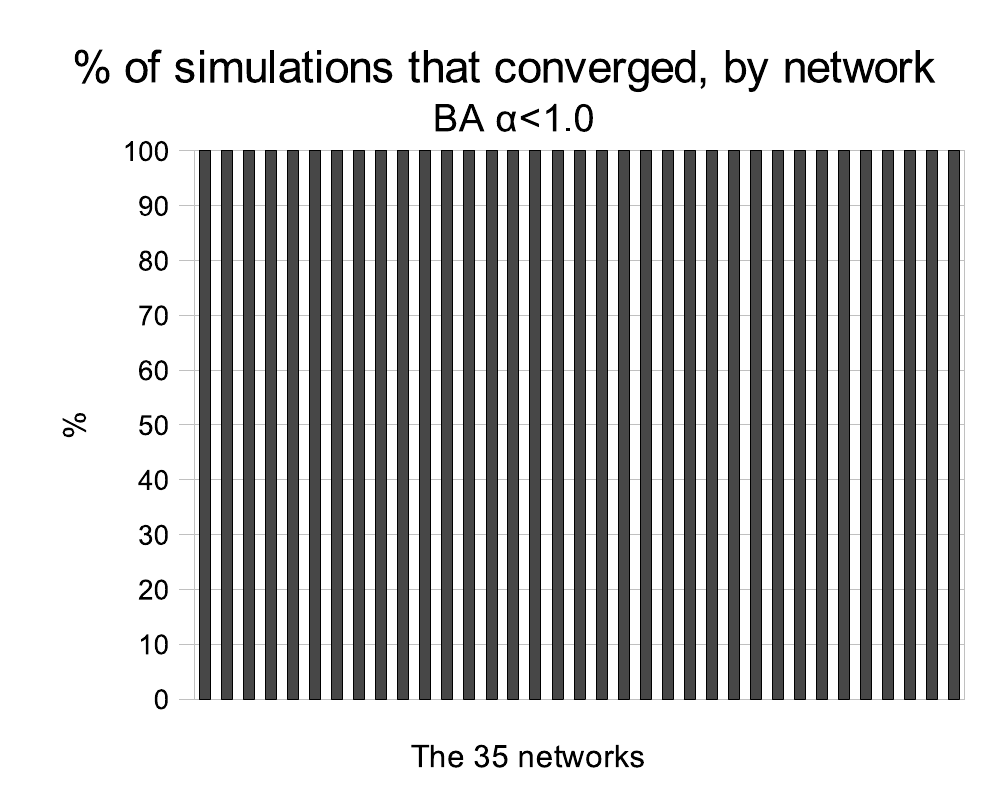}
\caption{Percentage of simulations that converged when $\alpha<1$ for
each of the 35 scale-free networks with at least one fixed point.}
\label{fig:perc_conv_ba_lt1}
\end{minipage}
\hspace{0.3cm}
\begin{minipage}[t]{0.45\linewidth}
\centering
\includegraphics[height=55mm]{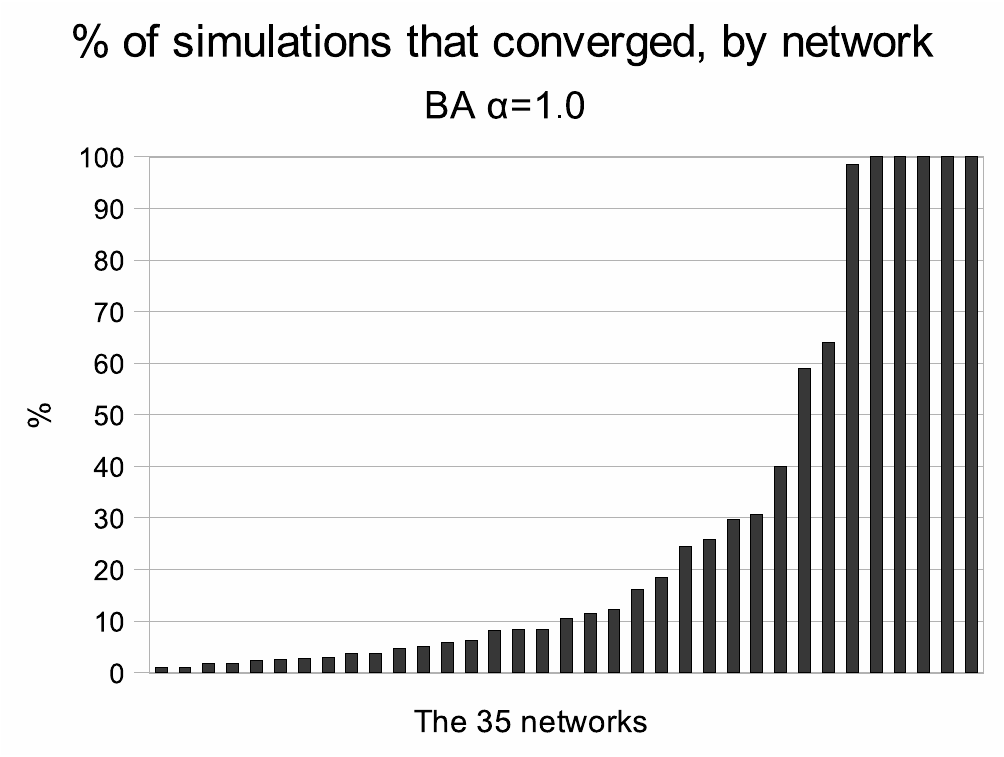}
\caption{Percentage of simulations that converged when $\alpha=1$ for
each of the 35 scale-free networks with at least one fixed point.}
\label{fig:perc_conv_ba_eq1}
\end{minipage}
\end{figure}

We begin by focusing our analysis on those 35 scale-free networks for
which we were able to prove the existence of fixed points (by finding
them). Recall that we used 1000 initial configuration per network. If
$\alpha<1$ (i.e, $\alpha=0,0.1,\ldots,0.9$) out of the 350000
stochastic trajectories we computed, 100\% of them reached a fixed
point. Despite the fact that for all these 35 networks we were also able to find
at least one converging trajectory  when $\alpha=1$, 
the situation in this case changed dramatically.
In fact, out of the 35000 deterministic trajectories we simulated for those
35 networks, only $\sim$28.9\% of them reached a fixed point. Notice
also that the  percentage of deterministic trajectories that reached a
fixed point varied greatly form one network to another. We present the
percentages for each network in Figure~\ref{fig:perc_conv_ba_eq1}. Every
single trajectory that did not reach a fixed point ended in a cycle
within the time horizon. 
For presentation purposes, we included the analogous diagram for  $\alpha<1$ 
in Figure~\ref{fig:perc_conv_ba_lt1}.

We immediately notice that a higher proportion of simulations reached a
fixed point when $\alpha<1$ than when $\alpha=1$. This is not
unexpected as the non-determinism in the stochastic trajectories is a
way to avoid the cyclic attractors the deterministic trajectories tend
to fall into. The remarkable property is that, if for a given network a
single trajectory with $\alpha<1$ converged to a fixed point, then
{\emph{all the trajectories}} with $\alpha<1$ also converged (although
not necessarily to the same point). Also note that all the fixed point
discovered through fully parallel updates were also discovered with the
$\alpha<1$ simulations. This indicates that stochastic updates can be
used as a robust method to detect the existence of fixed points.

\begin{figure}[ht]

\begin{minipage}[t]{0.45\linewidth}
\centering
\includegraphics[height=55mm]{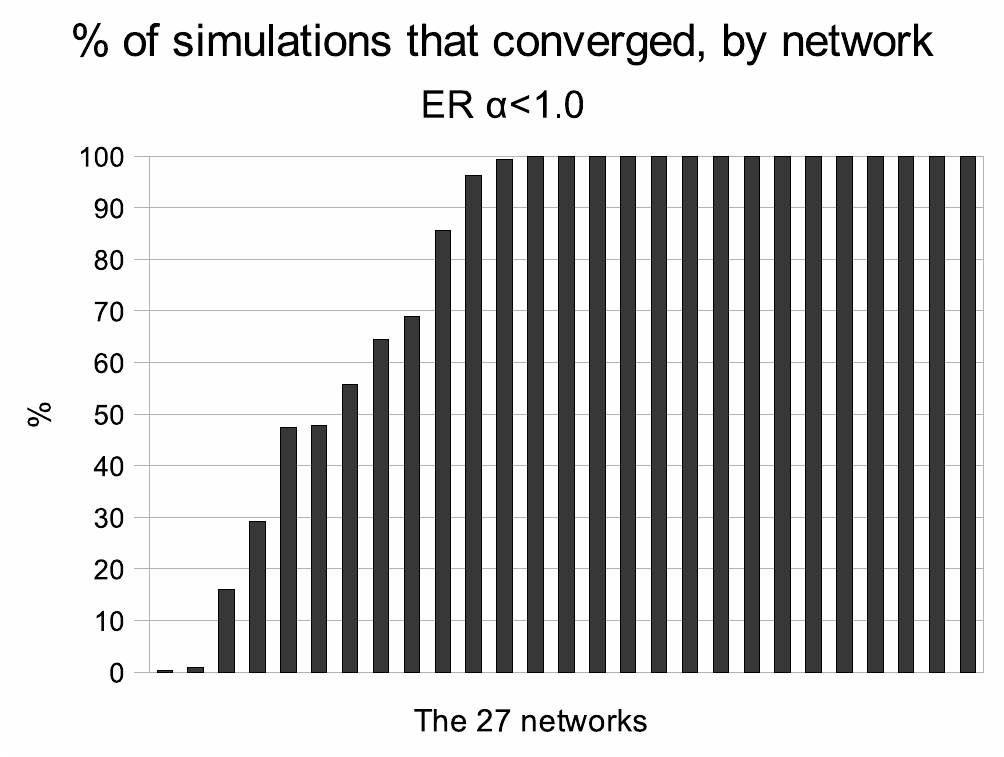}
\caption{Percentage of simulations that converged for each of the 27 ER networks where a fixed point was found.}
\label{fig:perc_conv_er_lt1}
\end{minipage}
\hspace{0.3cm}
\begin{minipage}[t]{0.45\linewidth}
\centering
\includegraphics[height=55mm]{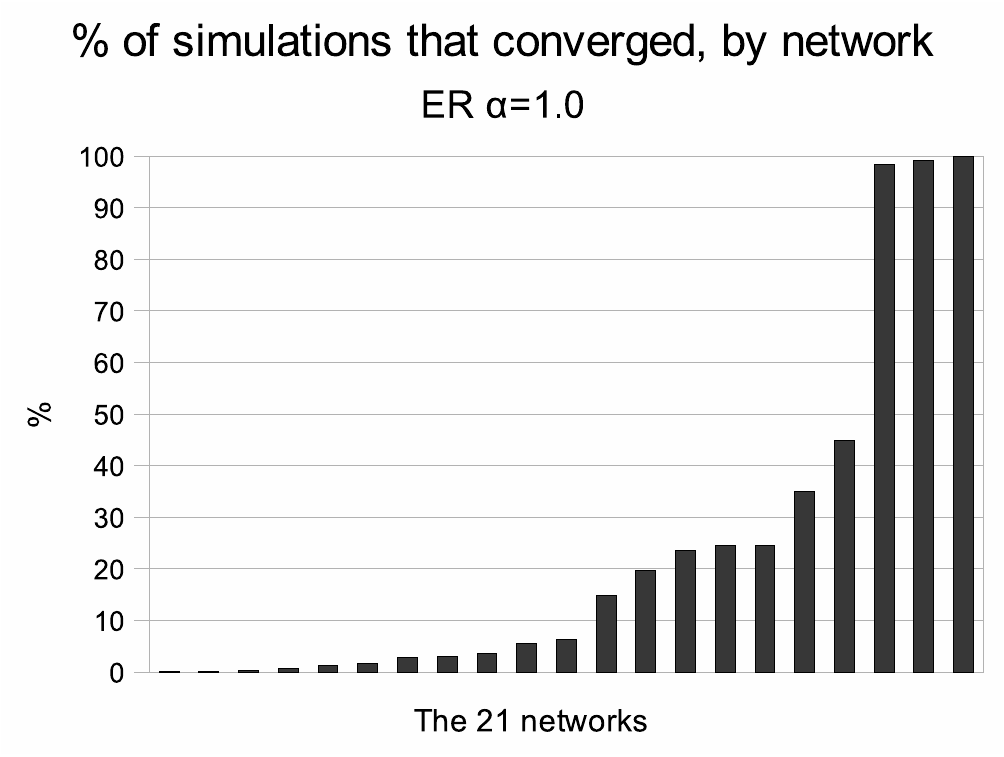}
\caption{Percentage of simulations that converged for each of the 21 ER networks where a fixed point was found.}
\label{fig:perc_conv_er_eq1}
\end{minipage}

\end{figure}

For the Erd\"{o}s-R\'enyi networks we computed the same statistics as we did for the
scale-free networks, were applicable. There were 27 out of the 50
networks were we found at least a fixed point. Interestingly, only $21$
of those fixed points were detected by deterministic trajectories. The
breakdown of percentages of converging trajectories by network for the
case $\alpha=1$ is presented in Figure~\ref{fig:perc_conv_er_eq1}.
Every single trajectory that did not reach a fixed point ended in a
cycle within the time horizon. The breakdown of percentages by network
for the case $\alpha<1$ is presented in
Figure~\ref{fig:perc_conv_er_lt1}. We notice immediately the difference
between Figures~\ref{fig:perc_conv_er_lt1}
and~\ref{fig:perc_conv_ba_lt1}: For some Erd\"{o}s-R\'enyi
networks with fixed point not all stochastic trajectories converged to a
fixed point while all trajectories converged in the scale-free networks
case, as we already described before.

\begin{figure}[ht]
\begin{minipage}[t]{0.45\linewidth}
\centering
\includegraphics[height=45mm]{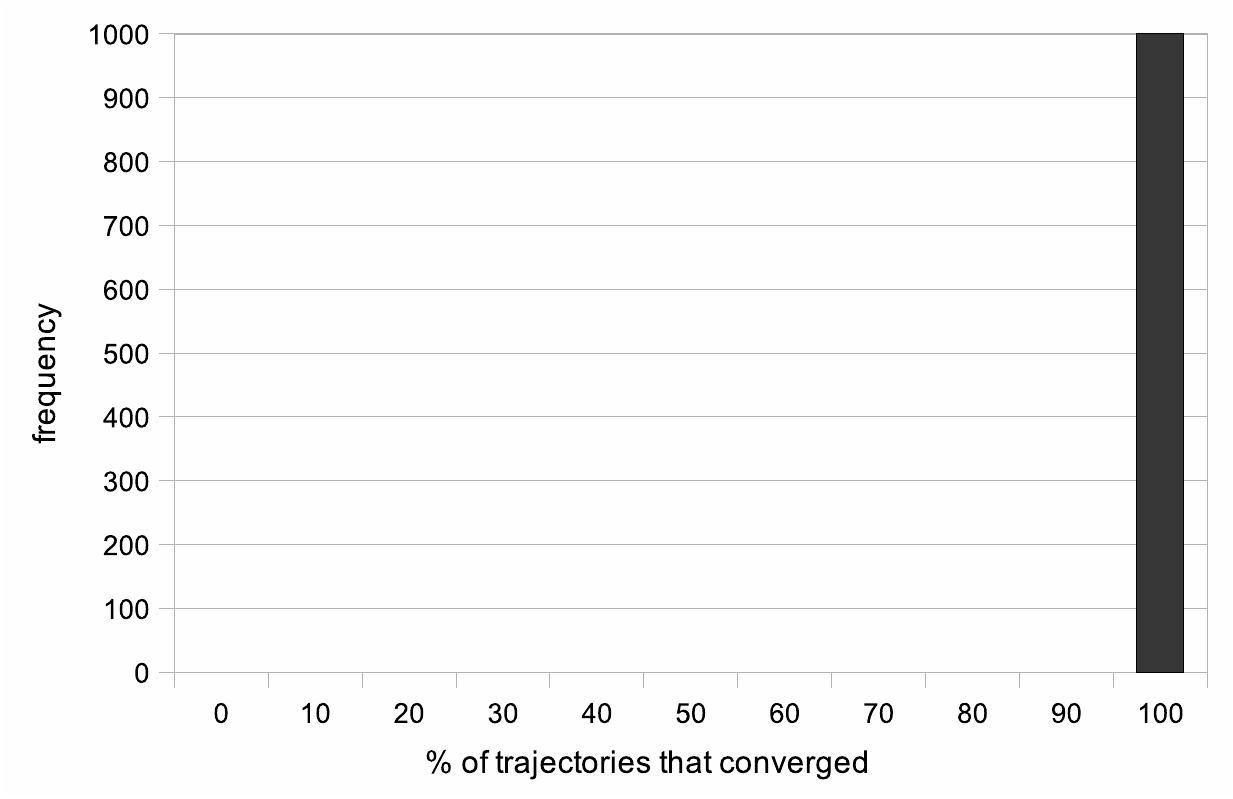}
\caption{Histogram of number of initial configurations that lead to a
converging trajectory for a single scale-free network which had at least one
fixed point ($\alpha<1$).}
\label{fig:ba_perc_convergence}
\end{minipage}
\hspace{0.3cm}
\begin{minipage}[t]{0.45\linewidth}
\centering
\includegraphics[height=45mm]{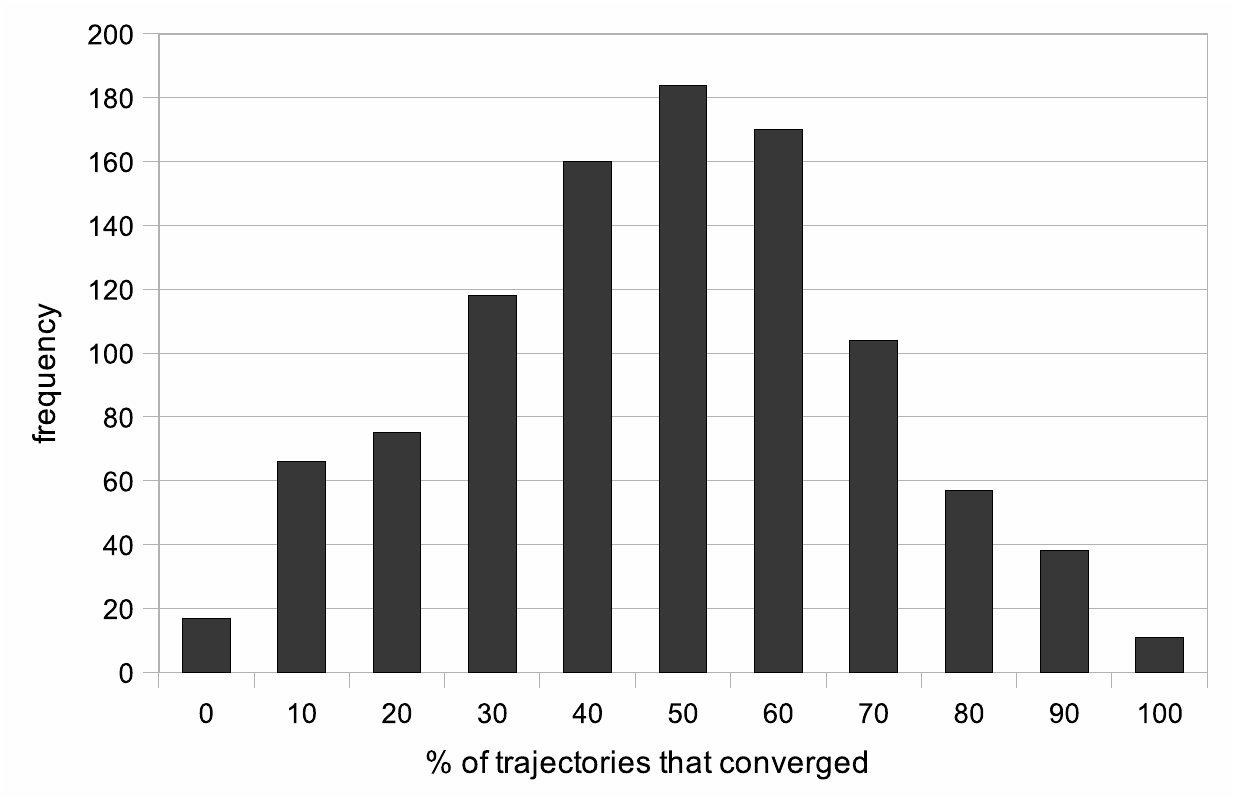}
\caption{Histogram of number of initial configurations that lead to a
converging trajectory for a single Erd\"{o}s-R\'enyi network which had at least  one fixed point ($\alpha<1$).}
\label{fig:er_perc_convergence}
\end{minipage}
\end{figure}

To emphasize the differences in the qualitative behavior 
we look more closely at two particular networks.
We selected one Erd\"{o}s-R\'enyi network for which the percentage of
trajectories that converged (considering all $\alpha<1$) was about 50\%. Recall
that 1000 initial configurations were used, 10 values of $\alpha$ were
tried and one trajectory per $\alpha$ was simulated. That means
that about 5000 out of 10000 trajectories converged to a fixed point for that
particular network. The histogram in
Figure~\ref{fig:er_perc_convergence} shows that the choice of the initial
configuration changes the probability of reaching a fixed point. The
heights of the bars represent numbers of initial configurations. The
horizontal axis is the percentage of trajectories (starting from one particular
initial configuration) that converged to a
fixed point. For instance,  for the Erd\"{o}s-R\'enyi network, we
can deduce the following: 160 initial configurations converged in 40\% of the simulations.
Since for each initial configuration we ran exactly 10
simulations there is no need to associate intervals with bins. For
contrast, we show in Figure~\ref{fig:ba_perc_convergence} the
analogous histogram for an arbitrary scale-free network with a fixed point.

\begin{figure}[ht]
\centering
\includegraphics[height=60mm]{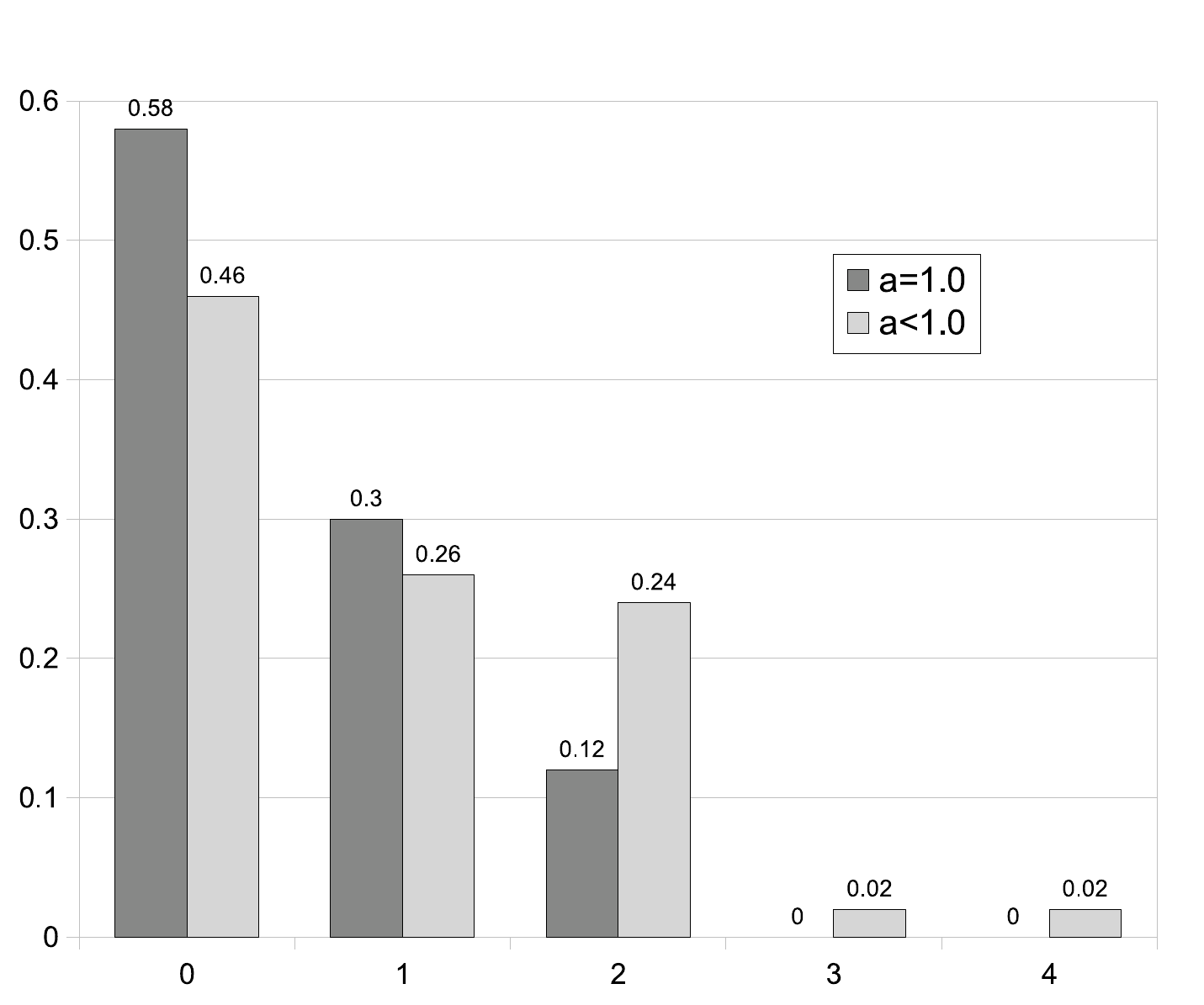}
\caption{Frequency of number of fixed points for ER networks for the
$\alpha=1$ and $\alpha<1$ cases.}
\label{fig:ba_effective_fixed_points_by_a}
\end{figure}

Another difference between the influence of parallelism when updating scale-free
or Erd\"{o}s-R\'enyi networks is seen comparing
Figures~\ref{fig:ba_effective_fixed_points},~\ref{fig:er_effective_fixed_points}
and~\ref{fig:ba_effective_fixed_points_by_a}.
Figure~\ref{fig:ba_effective_fixed_points_by_a} shows the difference in
number of fixed points found depending on $\alpha$ being 1 or less
than 1. The columns that describe the $\alpha<1$ case represent the
same values found in the histogram in
Figure~\ref{fig:er_effective_fixed_points}. The columns that describe
the $\alpha=1$ case in Figure~\ref{fig:ba_effective_fixed_points_by_a}
show that fewer fixed points were discovered. This, again, strongly
suggests that using partially parallel updates is a sensible strategy to
find fixed points. It also indicates that the set of discovered fixed
points somehow depends on the choice of $\alpha$. For scale-free networks, we
also discriminated between $\alpha=1$ and $\alpha<1$ cases. For these
networks though, in both cases the histograms were identical to the one
in Figure~\ref{fig:ba_effective_fixed_points}. This indicates that the
convergence to fixed points is fairly insensitive to the choice of
$\alpha$. This property shows a form of robustness of scale-free networks.

\subsection{Time and work until convergence}

\begin{figure}[ht]

\begin{minipage}[t]{0.45\linewidth}
\centering
\includegraphics[height=50mm]{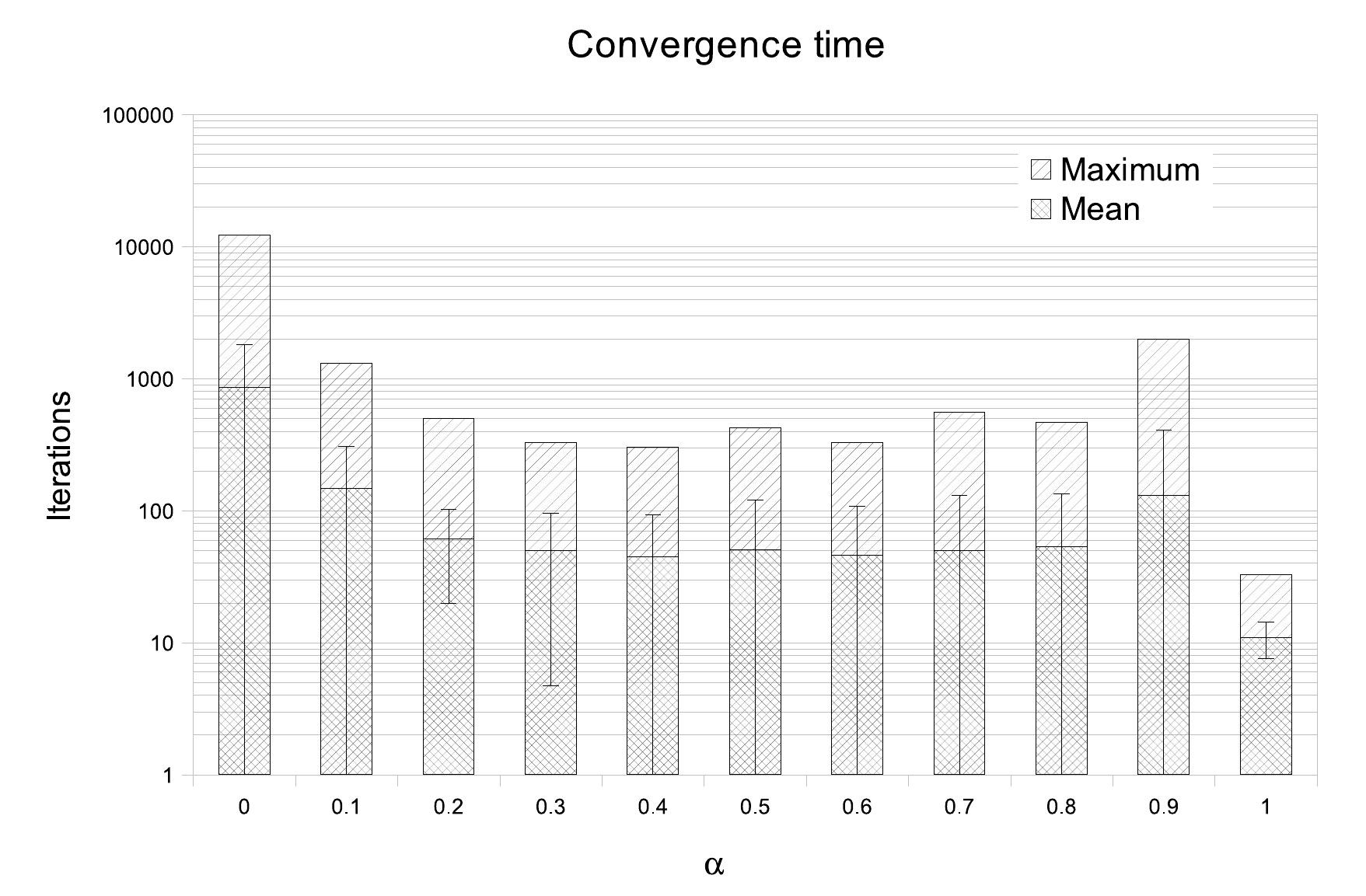}
\caption{Convergence time for scale-free networks}
\label{fig:times_ba_effective}
\end{minipage}
\hspace{0.3cm}
\begin{minipage}[t]{0.45\linewidth}
\centering
\includegraphics[height=50mm]{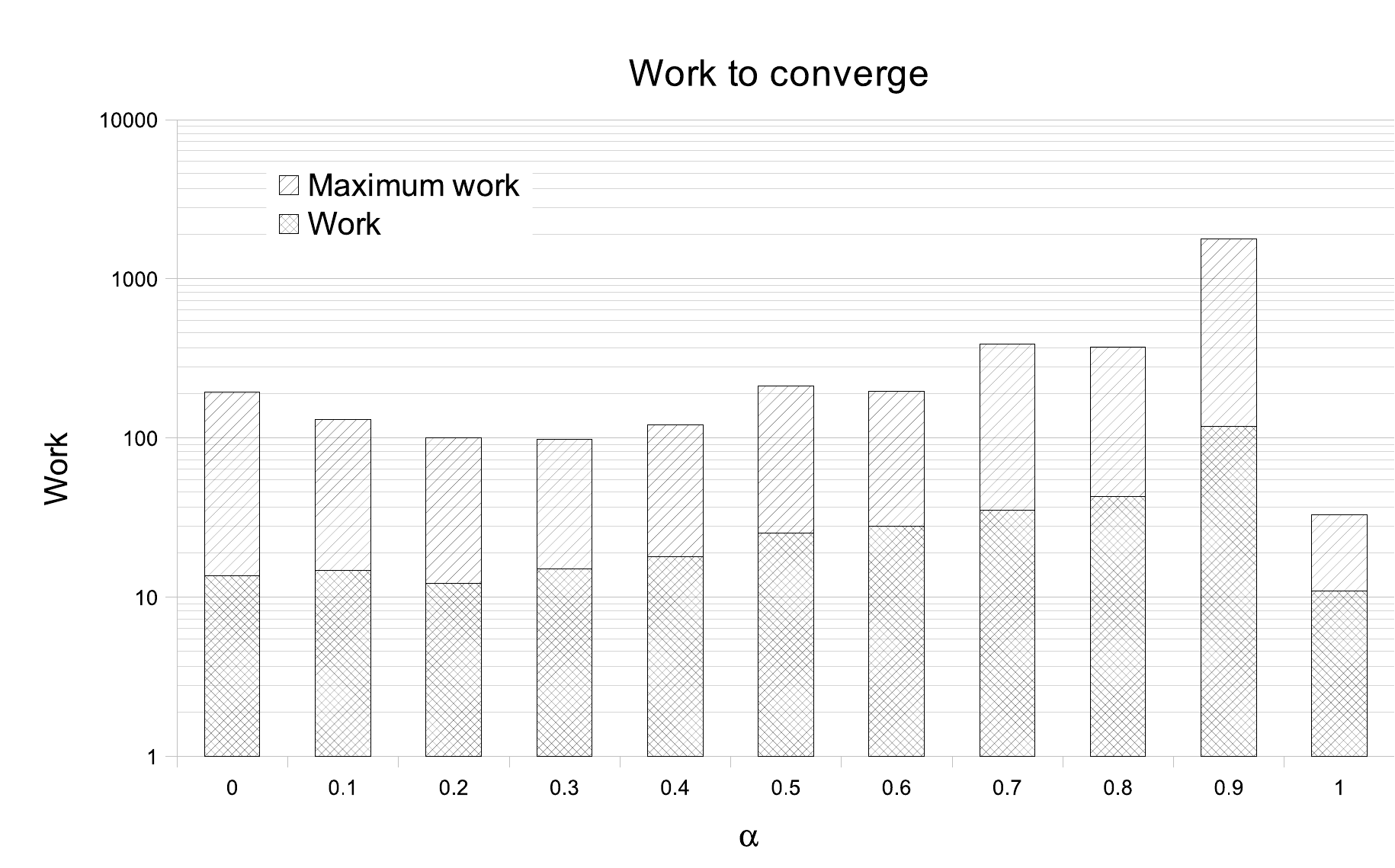}
\caption{Convergence work for scale-free networks}
\label{fig:work_ba_effective}
\end{minipage}

\end{figure}

For scale-free networks, Figures~\ref{fig:times_ba_effective}
and~\ref{fig:work_ba_effective} represent the mean and maximum time/work
to reach the fixed point for {\em only} the simulations that reached it.
This means, for $\alpha<1$, each column gives the average and maximum
over 35000 simulations, while for $\alpha=1$ the values were computed
over a smaller sample because only $\sim$28.9\% of the runs ended before
the time horizon was reached. This makes the values of the rightmost
column of each figure somewhat incomparable with the others.

In Figure~\ref{fig:times_ba_effective} we note how the average time
changes with $\alpha$. The qualitative behavior of the results are in
part intuitive. If $\alpha=0$ we expect the time to be high, since not
much work is done per iteration. We also expected the time to increase
when $\alpha$ approaches 1, because the system becomes more
deterministic and tends to imitate the behavior of the $\alpha=1$
case. Therefore, the cycles of the deterministic trajectories become
metastable regions of the stochastic trajectories.

\begin{figure}[ht]
\begin{minipage}[t]{0.45\linewidth}
\centering
\includegraphics[height=50mm]{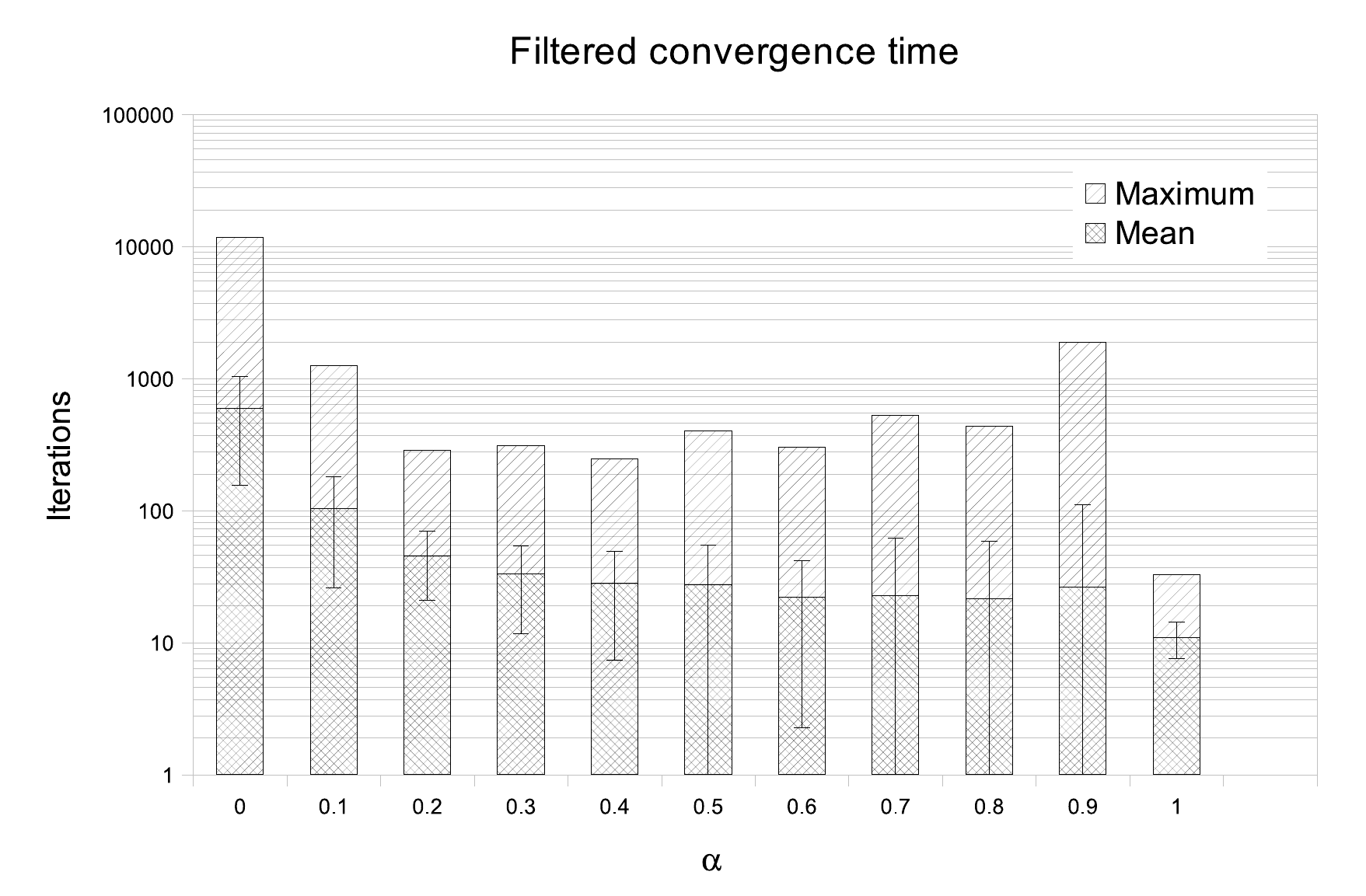}
\caption{Filtered convergence time for scale-free networks}
\label{fig:filtered_times_ba_effective}
\end{minipage}
\hspace{0.3cm}
\begin{minipage}[t]{0.45\linewidth}
\centering
\includegraphics[height=50mm]{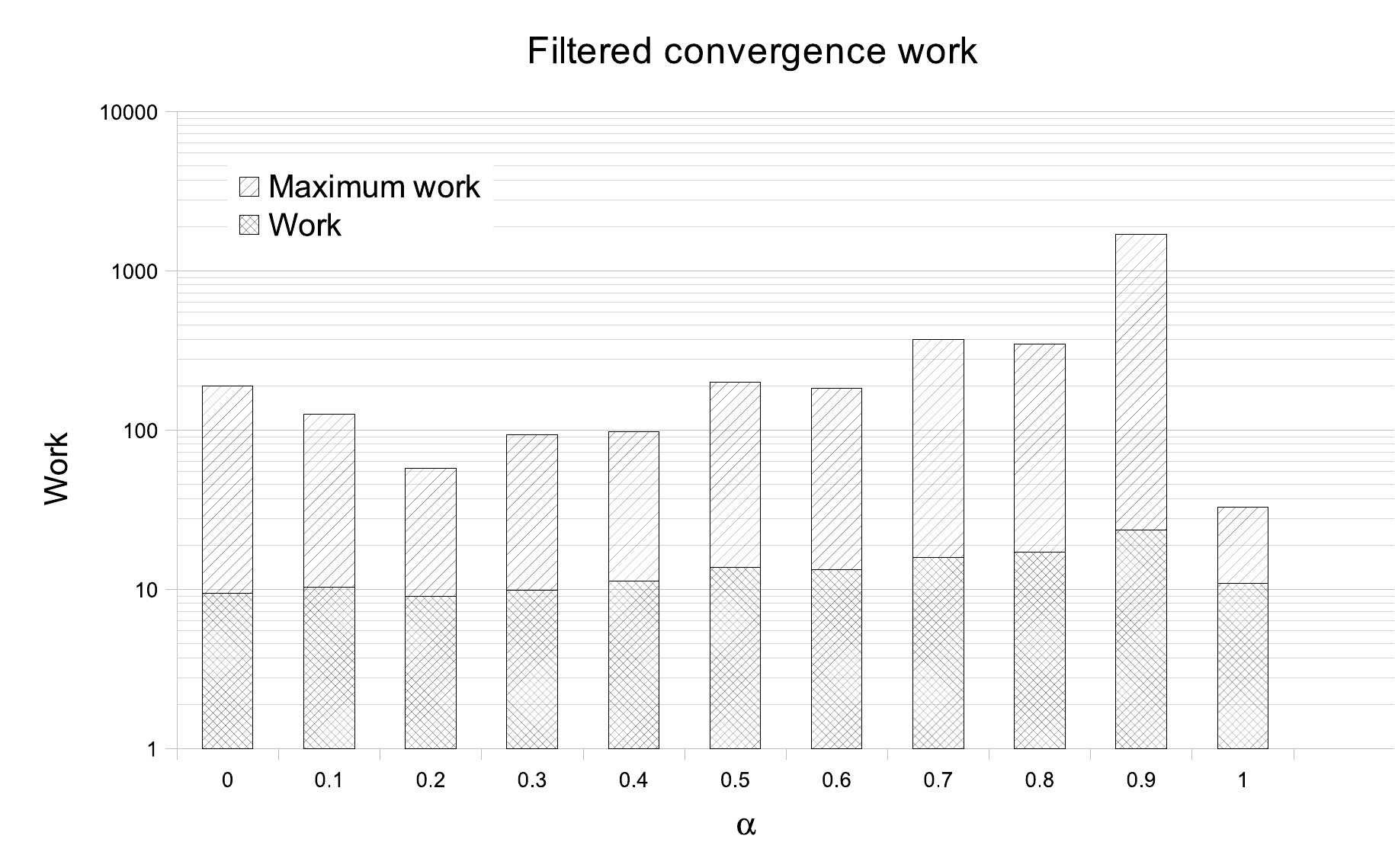}
\caption{Filtered convergence work for scale-free networks}
\label{fig:filtered_work_ba_effective}
\end{minipage}
\end{figure}


Note that the average number of iterations for $\alpha=1$ is smaller
than for all the values for $\alpha<1$. The comparison is not fair,
though, because the parallel update rule is not a reliable way to find
fixed points. With respect to the amount of  work,  we can notice, as
expected, that  small values of $\alpha$ become  very competitive (see
Figure~\ref{fig:work_ba_effective}). One possible explanation for the
parallel simulations outperforming the $\alpha<1$ cases (neglecting the
lack of robustness, of course) is as follows: The $\alpha=1$ updates are
prone to solve only the ``simple instances'' of the problem. That is,
only very stable fixed points located close to the initial configurations are
likely to be found.

To make a fair comparison between the $\alpha<1$ and $\alpha=1$
situations, we recompute the statistics but only for the cases where the
deterministic model found a fixed point. This means, we considered only
the pairs (network $G$, initial configuration  ${\mathbf s}$) such that
the deterministic trajectory of $G$ starting from ${\mathbf s}$ reached
a fixed point. Figures~\ref{fig:filtered_times_ba_effective}
and~\ref{fig:filtered_work_ba_effective} show the results. From
comparison against Figures~\ref{fig:times_ba_effective}
and~\ref{fig:work_ba_effective} we notice that the average times of the
simulations for $\alpha<1$ actually decreased. This supports the
hypothesis of $\alpha=1$ working mostly for simple instances of the
problem. The differences in running time may not be considered
substantial though and the problem requires further study.

\begin{figure}[ht]
\begin{minipage}[t]{0.45\linewidth}
\centering
\includegraphics[height=50mm]{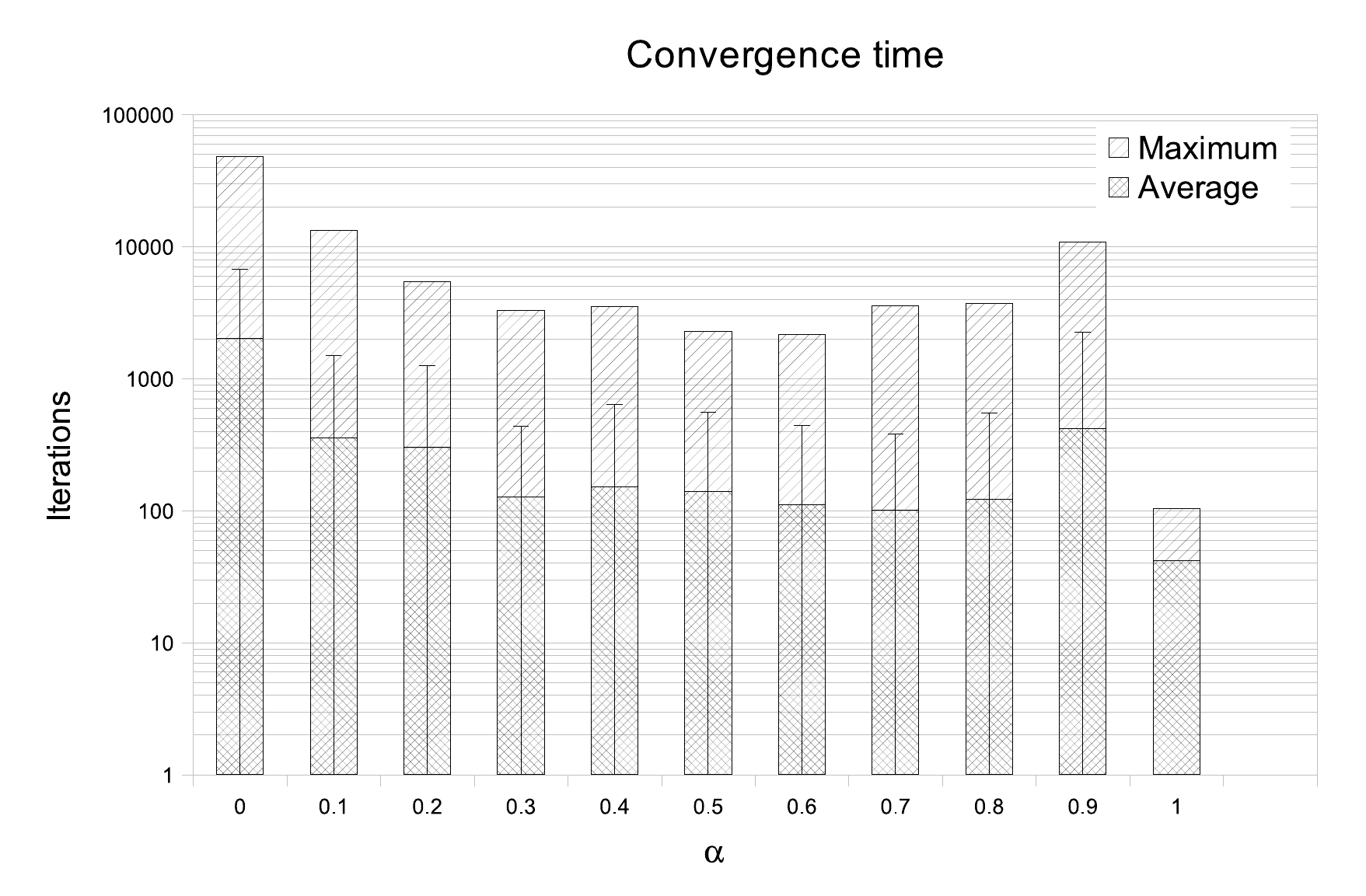}
\caption{Convergence time for Erd\"{o}s-R\'enyi networks}
\label{fig:times_er_effective}
\end{minipage}
\hspace{0.3cm}
\begin{minipage}[t]{0.45\linewidth}
\centering
\includegraphics[height=50mm]{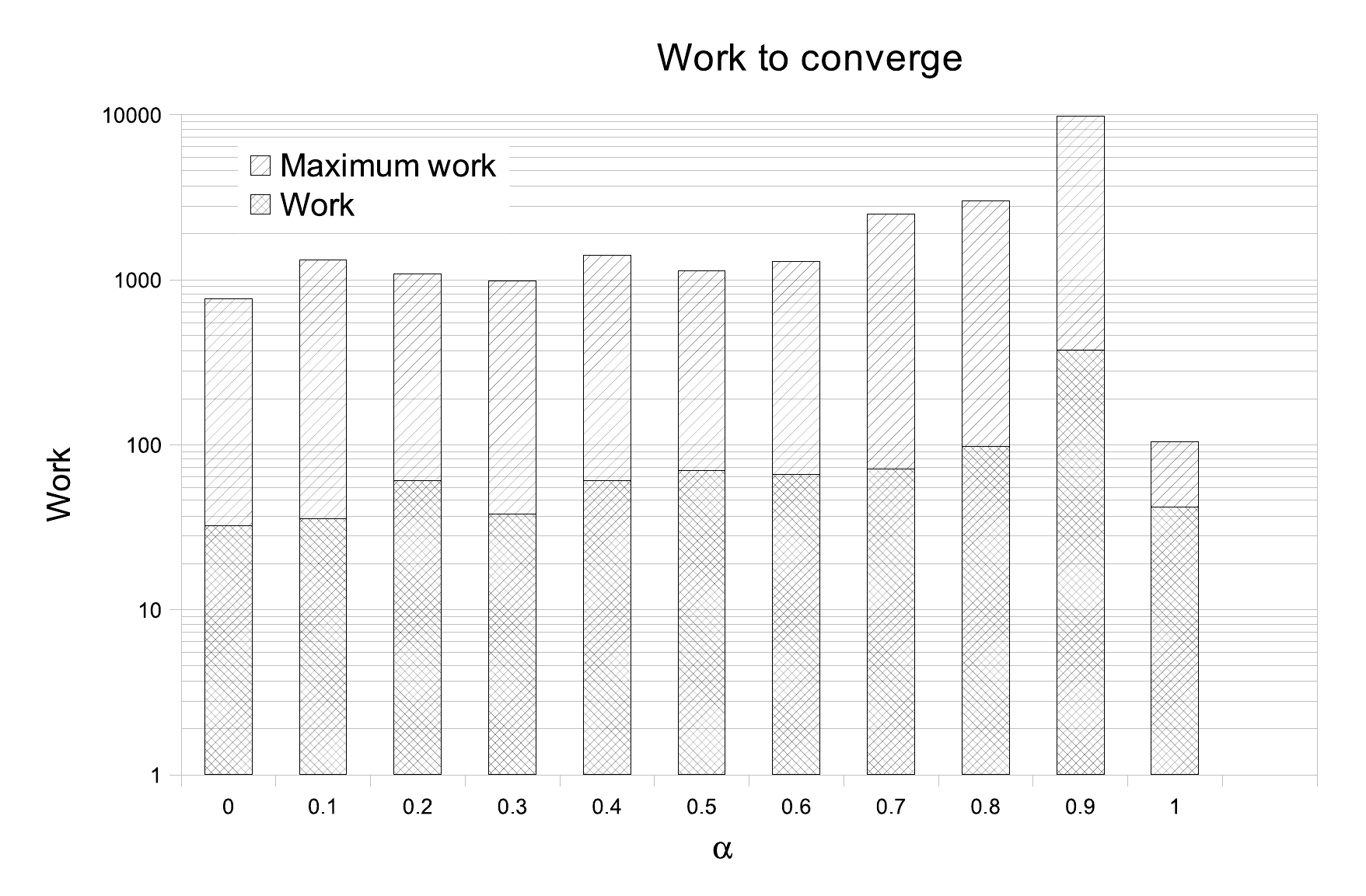}
\caption{Convergence work for Erd\"{o}s-R\'enyi networks}
\label{fig:work_er_effective}
\end{minipage}
\end{figure}

For Erd\"{o}s-R\'enyi networks, Figures~\ref{fig:times_er_effective}
and~\ref{fig:work_er_effective} are the mean and maximum times/work to
reach the fixed point for {\em only} the simulations that reached it.
This means, for $\alpha<1$, each column gives the average and maximum
over 27000 simulations, while for $\alpha=1$ the values were computed
over a smaller sample because only $\sim$24\% of the runs ended before
the time horizon was reached. In this case we did not compute the
averages/maximums for the stochastic model restricting the networks and
initial conditions to those that reached a fixed point with
deterministic updates. There was no qualitative difference between the
results for scale-free networks (Figures~\ref{fig:times_ba_effective}
and~\ref{fig:work_ba_effective}) and the times for Erd\"{o}s-R\'enyi networks. Data
suggests that Erd\"{o}s-R\'enyi networks are slower to converge than scale-free systems.

\subsection{Discussion}

We will first make the case that the RBNs, as defined here, have a
behavior that resembles that of natural networks. It is well known that
gene expression/repression changes according to external cell
conditions, such as temperature or availability of nutrients and/or
oxygen. For brevity, we will refer to such conditions as \emph{EC}. This
can be modeled easily by setting the values of input nodes  depending on
the EC. It is also widely accepted that gene expression state is (after
the transient) {\it mostly} function of the EC. In terms of the model
this means the fixed point should be function of the values of the input
nodes. Although stochasticity sometimes plays an important role in cell
dynamics~\cite{Kaern2005}, for purposes of discussion we will assume
that for each set of EC there should be one or a small number of fixed
points. Another property is stability, meaning the fixed point should
have a large basin of attraction. Additionally, the fixed point should
be reached ``quickly.'' By quickly, we mean that the number of
transitions should be small compared to the total number of possible
configurations. Finally, the behavior of the model should not depend
dramatically on the choice of $\alpha$.

Our simulations with asynchronous updates on scale-free networks showed
all properties mentioned above. The theorem in
Subsection~\ref{subsec:dynamics}, together with the histogram in
Figure~\ref{fig:ba_effective_fixed_points} address the number of fixed
points is, with high probability, close to one. The fact that all
trajectories converged, as shown in Figure~\ref{fig:perc_conv_ba_lt1}
shows the model is (robust/stable?) regardless of the choice of
$\alpha$, as long as it is less than $1$. With respect to the speed of
convergence, the average number of iterations to reach a fixed-point
(see Figure) is small compared to the number of possible configurations,
which was about $2^{63}$ in all the numerical experiments.

Of course, the goal of the model is to provide insight on biological
process. The most remarkable observation is that the three
aforementioned properties appeared in the simulations without requiring
any kind of deliberate design. The network topologies, the dynamics and
vertices to be updates were selected at random. Yet, the behavior was as
desired, \emph{as long as the network was scale-free and $\alpha<1$}.
This suggests that the prevalent scale-free topologies in natural
GRNs~\cite{Albert2005} could be a major factor in
robustness of living cells.

As an aside, finding fixed points of a Boolean network is equivalent to
finding satisfying assignments to Boolean formulas. This is a well known
problem in Computer Science, and is believed to be computationally hard
if we consider the worst case running time. However, the results shown
in Figures~\ref{fig:work_ba_effective},
\ref{fig:filtered_work_ba_effective} and~\ref{fig:work_er_effective},
which implies fast convergence, suggest that the asynchronous update
rule is an efficient heuristic to find fixed points or satisfying
assignments.

\section{Conclusions}

We analyzed the tendency to reach a fixed point, the number of
iterations and the work needed to converge to it for different families
of Boolean networks and update policies. We summarize our results as
follows:

\begin{enumerate}

\item Using partial parallelism (i.e, $\alpha<1$), the likehood of the
trajectory to reach a fixed point increases. This happens regardless of
topology.

\item For scale-free networks the choice of $\alpha$ is
not very important from the convergence point of view, as long as it is
less than one. We base this conclusion on the fact that we could not
find a simple example of a network not converging to a fixed point
provided that it has 1.

\item For scale-free networks  the choice of $\alpha$ is
not very important from the work  point of view, as long as it is less
than one. This follows from the analysis of the filtered case. The work
is about the same as in the $\alpha=1$ case, while full parallelism
increases the likehood of being trapped in a cycle.

\item For Erd\"{o}s-R\'enyi networks the choice of $\alpha$ is  relevant from
the work point of view, when it is less than 1. The set of points
discovered seems to be dependent somehow on the choice of $\alpha$.

\item The combination of scale-free networks and using $\alpha<1$ for
the updates was enough to impose dynamical properties which are similar
to those observed in nature.

\end{enumerate}

\section{Acknowledgments}
We would like to thank \'{A}lvaro Olivera for useful discussion.

\end{document}